# Designing Technology for Positive Solitude

Pertti Saariluoma[1], Juhani Heinilä[2], Erkki Kuisma[3],
Jaana Leikas[2], Hannu Vilpponen[3], Mari Ylikauppila[2]

[1]University of Jyväskylä; firstname.lastname@jyu.fi
[2]VTT Technical Research Centre of Finland; firstname.lastname@vtt.fi
[3]Nokia; firstname.lastname@nokia.com

*"When men are merely submerged in a mass of impersonal human beings pushed around by automatic forces, they lose their true humanity, their integrity, their ability to love, their capacity for self-determination. When society is made up of men who know no interior solitude it can no longer be held together by love: and consequently it is held together by a violent and abusive authority. But when men are violently deprived of the solitude and freedom which are their due, the society in which they live becomes putrid, it festers with servility, resentment and hate." (Thomas Merton, Trappist monk) (Merton 1958).*

**Abstract:** This paper discusses Life-Based Design methodology (LBD) within the context of designing technologies for reaching a state of solitude, i.e., the state where a person wishes to minimize her social contacts to get 'space' or 'freedom'. LBD is a multi-dimensional approach, which emphasises the importance of *designing for life*, i.e. understanding people's lives (forms of life and circumstances) to derive design ideas and to carry out concept design. In this article, we discuss solitude in relation to the emerging need of rethinking the realization of information society and the desire of 'disconnecting' ourselves from the hectic 24/7 online life. We follow the Life-Based Design process in developing a concept for solitude service, i.e., reaching a state of solitude supported by electronic gadgets.

*Introduction*

Designing new technologies presupposes understanding of everyday life, as all technologies are justified by their capacity to improve the quality of life. However, methods for innovation and development of products and services on the ground of information on human life are still under-researched and underdeveloped. Before adequate technologies can be designed there has to be a clear understanding of the users' needs, that is *what* the new technologies will be designed *for* and how they are to be incorporated to the everyday life. Without a clear idea of design grounds, users' skills and human goals, it is difficult to realize new technical ideas. Consequently, one should inquire about concepts that would allow for the best understanding of the properties of human life, and find out which of these concepts could be systematically used in developing technologies. Finally, the ideal process for undertaking this analysis of life and the design for it should be considered.

Life-Based Design (LBD) is a design methodology which emphasizes the importance of the understanding of human life and life science as a basis of early-stage design (Leikas 2009; Saariluoma and Leikas 2010). It includes processes which enable designers to derive goals for technology design through analysing human life and thus provides a powerful tool for conceptualization based on everyday situations and solutions to everyday problems. The main task of LBD is to analyse human life for technology design. The concept *form of life* is here used as the ground concept.

The form-of-life notion is originally from Ludwig Wittgenstein's (1958) philosophy of language, and by it in LBD is referred to any system of regular actions in human life. Human actions are controlled by different regularities, facts and values in life. For example, people may be students or unemployed, civil servants or farmers, football fans or keen golfers, and when following these roles they participate in a specific form of life. The number of forms of life is unlimited. When designers are able to understand the basic structure and logic of some form of life, they are then provided with information to assess what kinds of technologies might serve this particular form of life.

When analysing forms of life, one can start by searching for solutions to the basic issues regarding what kinds of socio-cultural contexts people participate in, and what are their actions relevant to these contexts, or what kinds of biological and psychological facts explain the reasons for their behaviour (Leikas and Saariluoma 2008; Leikas 2009; Saariluoma and Leikas 2010). To understand the internal structure of some form of life, it is essential to understand what are the regular action patterns, facts and values that people have in order to accept that particular form of life and be able to participate in it. The goal of the analysis of life and that of a particular form of life is to open up these fact and values.

A typical fact of life is the situation in life. From the point of view of situation in life, there are additional concepts to consider. The circumstances of a situation in life can be determined by, e.g., illness, wealth or poverty, youth or old age, being a student or an enterprise leader. Such situations are facts of life, and they make it understandable as to what kinds of lives people live. The core idea of Life-Based Design is that the knowledge about the structure of the forms of life can be used when designing technologies such as ICT-services for improving the quality of life. In this way, we call attention to the fact that designing for life is focused on opening new possibilities instead of merely solving problems.

Sharing a form of life does not mean that people would be in direct contact with each other, but rather like an audience of a soccer game: some of them may be at the stadium watching the game, some in front of TV-screens and some reading about the game in local newspapers. Nevertheless, their situation of life is similar in the sense that they all follow soccer. Thus, when many people have the same form of life or the same life situation, it makes sense to develop services for them. Consequently, Life-Based Design can be grounded on the analysis of life situations shared by a large number of people.

In this paper, we illustrate a concrete case to apply Life-Based Design in order to outline human goals. We have specifically focused on a specific type of form of life, which is determined by a mental state people hope to reach. This state is solitude. Of course, it could also be something else, such as social discourse, finding friends, or immersing in a game. Our goal is to illustrate a way of design thinking when designing for life, and in this purpose solitude serves as a good example of a situation in life determined by a mental state people hope to reach.

### *What is solitude?*

Solitude refers to 'aloneness', but is often a consequence of a voluntary decision and associated with forms of life of the kind where a person may minimize her social contacts to get 'space' or 'freedom'. Many poets, writers and artists have experienced solitude as a regular part of their creating process. Storr (2005) argues that the accomplishments of many creative persons, such as writers, artists, composers, and philosophers, would have been impossible without their precious time alone to think things through.

In particular, the act of being alone has been understood as an essential dimension of religious experience in different religions, and many religious persons have actively sought solitude for meditation and inspiration (Deresiewicz 2009; Long, More and Averill 2006). However, solitude should not be misunderstood as either a state of loneliness or meditation. Loneliness is involuntary, and therefore it is unpleasant. Meditation, on the other hand, is work and has to be performed in solitude. Solitude is a primary condition for starting meditation. According to Andre (1993) solitude provides the opportunity to identify your most cherished goals and develop ways of achieving them. Regular reflection contributes to a sense of inner peace and makes you feel more in control of your life.

A state of solitude is loneliness when you long for someone. When there is a sensation of being isolated or ignored persisting for a longer duration, it can turn into a negative feeling. When solitude is experienced negatively, the person is surrounded by feelings of loneliness and depression. In a study by Long, More and Averill (2006), both positive and negative solitude as mental states were found to occur frequently. Both states tended to occur while the person was alone, both were preceded by a sense of stress, and both were likely to occur in environments close to home. Women were more likely to experience solitude at home, while men were more likely to feel it outdoors (Long, More and Averill 2006). Bailwal (2011) argues that loneliness, if used positively, is solitude. If we can drive energy and inspiration from loneliness, then it is not loneliness but solitude.

Furthermore, it is important to distinguish solitude from the concepts of 'privacy', 'well-being' and 'individualism' so that we have a clear understanding of what this phenomenon of solitude and situation of life is like. *Privacy* has meant physical isolation from other people, and it is thus close to solitude. However, in today's ICT-world, privacy refers to the right and possibility to prevent personal information to be divulged to unwanted people. This is different from solitude, which refers to (a temporary) absence of social relations (Burger 1995). Positive solitude is also connected to *well-being* in a positive sense. Well-being differentiates solitude from loneliness, which is characterized by negative attributes such as being isolated and ignored. *Individualism* refers to being self-centred in social contacts rather than to withdrawing oneself from contacts with other people. Solitude thus means capacity to do well in isolation from other people (Holenhorts, Frank and Watson 1994) It is often characterized by such positive attributes as *positivity, wholeness, finding oneself*, and even *solemnity* (Burger 1995).

Solitude enables us to secure the integrity of the self as well as to explore it. No real excellence, personal or social, artistic, philosophical, scientific or moral, can arise without solitude (Deresiewicz 2009). The search for solitude, as the basic concept in our example for designing services on the ground of the situation of life, is chosen here for several reasons. Firstly, solitude is really a situation of life. Secondly, as we understand it, it is a positive situation. Moreover, solitude can be seen as a pursuit towards a certain emotional situation in life. Therefore, we can classify it as a form of life just as we can classify, e.g., sports enthusiasm also as a form of life.

*Means for reaching a state of solitude*

Enjoying solitude frequently can help in solving a number of problems in our life and provide benefits, which we simply loose every day because of our extremely complex lifestyle (Bailwal 2011). There are numerous ways of reaching a state of solitude. In literature, we can find out that such actions as wandering in nature or going to church may equally provide for an individual an opportunity to a state of solitude. The main criterion is that a person can concentrate on one's thoughts and inner life and become separated from everyday social life. Places can be important in that they can separate one from the ordinary flow of life.

Hammitt and Brown (1984) identified five dimensions of wilderness privacy among student backpackers: emotional release, personal autonomy, reflective thought, limited communication: personal distance, and limited communication: intimacy. Later Hammitt and Madden (1989) extended the initial study and identified five factors important in wilderness privacy: tranquillity and natural environment, individual cognitive freedom, social cognitive freedom, intimacy, and individualism. From these results, Hammitt and Madden concluded that the essence of wilderness privacy involved being free of human-generated intrusions in a remote natural environment, where people had freedom over their own time and actions, as well as control over everyday pressures and attention loads. (Long, More and Averill 2006.)

According to The Chronicle Review, technology is taking away our privacy and our concentration, but it is also taking away our ability to be alone (Deresiewicz 2009). One can fairly ask whether it is possible to use modern technology for achieving solitude. Where and how could we find a solution to prevent the further loss of positive solitude? In this paper, novel ways to become 'disconnected' – i.e., reaching a state of solitude supported by electronic gadgets - will be studied, following the LBD methodology.

*The practise of Life-Based Design*

Life-Based Design as a method defines a procedure for innovating and designing human-technology interaction concepts. The procedure for Life-Based Design was described and reported in detail recently by Leikas and Saariluoma (Leikas 2009; Saariluoma and Leikas 2010; Leikas, Saariluoma, Heinilä and Ylikauppila 2013). Life-Based Design methodology consists of four different phases, which guide the designers' thinking along the design process. These phases are: 1) Form-of-life analysis, 2) Concept design and design requirements, 3) Fit-for-life design, and 4) Innovation design (Figure 1).

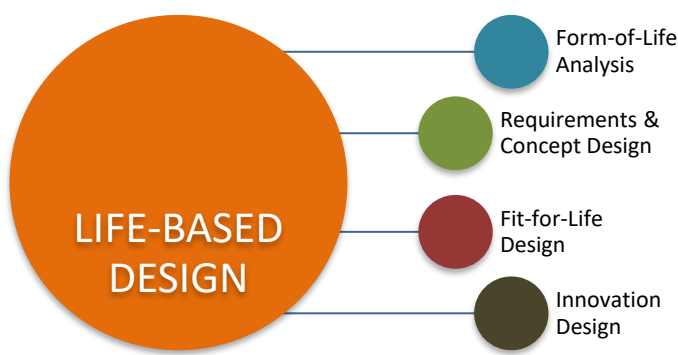

*Figure. 1. Life-Based Design framework (Leikas et. al. 2013).*

*Form-of-life analysis* begins with an analysis of the particular form of life the designers are interested in. To gather design-relevant information about the selected form of life, we have to extract the major regularities and other properties of that form of life. This means that we have to look at the kinds of actions people normally follow when they participate in it. It is important to see what the major explanatory facts and the life settings are and what kinds of value structures one can find in that particular form of life. As an example, this may be 'living in a senior home'. The general goal is to get a clear idea about what the properties (rule-following actions) of the particular form of life are. The steps to investigate this include:

- Analysis of the selected form of life;
- Definition of design goals;
    *What people need in their life and how technologies could really improve their life;*
- Explication of design-relevant problems;
    *Extraction of all the design-relevant human-based problems to define possible problematic side issues and put them under scrutiny;*
- Analysis of typical actors;
    *A realistic understanding about the potential users or actors and their properties, such as education, age, gender or technology skills;*
- Analysis of contexts;
    *Including both physical and social conditions and social relations activated before, after and when using the technology;*
- Analysis of other relevant characteristic actions.

The analysis of the form of life should generate the *human requirements* for a product or service. This is information explaining the *Why's* and *What for's,* which should guide the design process from the beginning to the very end. The human requirements define how people's life in a specific form of life should be improved. The human requirements are based on the methods and results of human life sciences and create the basis for the next phase in the design by introducing the design theme and the human requirements behind it. However, they do not yet define the requirements for technological concepts, which could be used in answering to the defined design goals of the specific form of life. These, in turn, are called 'design (user) requirements' and are illustrated in the next chapter.

*Concept Design and Design (User) Requirements* is the second phase of LBD. In this phase, the designers will define the role of technology in achieving the defined design goals and produce a definition of technology-supported actions in a product or service concept. This means defining what the technology is used for. The outcome of this phase is a definition of technology-supported actions in a product or service concept. In this phase, the designers also begin to outline what the supposed technologies could be like. They generate prototypes of the relevant new technologies. They also explain how, by defining the forms and role of technologies in the form of life, these technologies can be associated to the problems, and how these technologies can be implemented. Thus, the actual user and technical requirements are defined in this second phase.

Congruent with the ISO standard (ISO 1998, 1999, 2000, 2008), this concept-design phase carries out user interface design as well as usability and user experience studies. This may include, e.g., deciding between the use fixed or mobile technology or general purpose and special devices for realizing the particular design goal. Consequently, this includes the traditional usability engineering processes as a part of Life-Based Design (Leikas 2009) together with proper knowledge of potential users and use contexts.

The concept-design phase ends up with technical design and implementation. Here the designers generate mock-ups and prototypes by using the kind of technologies that should be able to reach the design goals and make it possible for the users to reach their life goals. To summarize, this phase of LBD includes:

- Definition of the role of technology in achieving action goals;
- Solution ideation and reflection;
- Elaboration of selected solutions;
- UI design;
- User evaluations (usability);
- Technical design and implementation.

The final outcome of this phase is a definition of the technological concept and how people shall use it in their lives. This includes (user-)requirements specification for the implementation.

*Fit-for-life design* is the third phase of LBD. It refers to examining the benefit and meaningfulness the users can get from the developed solutions and the impact the solutions have to the quality of life. The outcome of this analysis can lead to improvements and modifications of the product ideas. This is the most fundamental phase in the LBD framework, as it illustrates the logic to enhance the quality of life. It stresses the importance of coming back to the human requirements defined in the first phase of the process and reflecting upon them as well as defining and removing the barriers for implementation. Should there be any problems in fitting the outcome to the form of life, the prototype should be refined in accordance to the requirements. Only in this way it is possible to assure that the outcome really satisfies the needs of the users.

Furthermore, it is essential also to bear in mind the ethical considerations with respect of the concept. Ethical evaluation of the outcome is thus a natural part of this phase (Bowen 2009;Hongladarom and Ess 2997; Winston and Edelbach 2006). As ethics defines what can be considered as 'good of man', ethical analysis may help in exploring from whose perspective and by what kind of choices it could be possible to generate an increase of goodness and develop products with higher value in improving the quality of life.

To summarize, the essential parts of this phase are:

- Illustration of the logic to enhance the quality of life;
- Fit-for-life evaluation;
- Ethical evaluation.

*Innovation Design* is the final phase of LBD. The purpose of this phase is to introduce a procedure for exporting the design outcome into general use and incorporating the new technology in human life settings. This process entails activities which can transform the concepts into products used in everyday life. Good design is a prerequisite to ensure that the future product is really used by people in their life, and therefore innovation diffusion processes are a vital part of Life-Based Design.

To make something an innovation, it is important to create usage cultures for design outcomes. To ensure that people really get from the technologies what they wish to get, calls for guiding and teaching. It may also call for complicated educational programs. Without skilled users there can be no use cultures. A use culture also means that users have a role for technology in their life. Thus, the users should know how the quality of their lives could be improved with the help of the given new technologies.

Innovation design consists of definitions of the infrastructure, a marketing plan, and service and auxiliary activity. These are methods of conveying the message about the benefits of using the technologies to the users. The outcome of the innovation design process is a product, which has found its users, or at least a clear plan of how this goal can be reached. Thus, this final phase defines a procedure for exporting the outcome into general use.

## *Analysis of solitude as a form of life*

Deriving plans for technology on the ground of conceptual intuitions is not necessarily rational. It is necessary to have a more concrete idea about the phenomenon in its complexity if one wishes to design good technologies serving human need to solitude. This presupposes empirical work on the properties of solitude as it is understood today and as it has been seen earlier. This is also why it is vital to collect information about how potential users experience and value solitude.

In our study, the form-of-life analysis was carried out by using a storytelling method among 29 people of the age 20-55 (average age: 44). The participants were mainly people with academic education, 19 females and 10 males. Of them 8 lived in a single person's household and the rest of them in households of two or more people. As discussed above, most of human beings seem to long for some state of solitude occasionally. Hence, as for *definition of the typical actors,* the participants in this case study were both a homogeneous (e.g., education) and a heterogeneous (e.g., age, family size) group of people. It is worthwhile to notice that this group of people is not the only one that might long for solitude.

The participants providing the narratives were activated by a presentation of a few relevant and open questions dealing with solitude, as follows:

- Which ways of being alone in a state of solitude do they have?
- What are their positive states of solitude like? How have they experienced positive solitude?
- What are the most essential things, artefacts or issues used by them to reach a state of positive solitude?
- What are the disturbing factors from their positive solitude point-of-view?

According to the narratives, the busy life style with tight time schedules and varying roles (e.g., profession and career, being a mother/father) of modern people was often experienced stressing and problematic. Therefore, it might be asked how this situation of life could be lightened, or what should be done to find ways out from such situations of life.

Hence, the *definition of the design* problem is one of the first subtasks of form-of-life analysis, leading to *the design goal*, i.e., how technology could be harnessed in supporting people to achieve solitude in their lives.

Two important phases of the form-of-life analysis are the *identification of relevant design related actions* people are using for reaching their goals (e.g., getting into a state of solitude followed by the *typical contexts* where such actions take place. According to the narratives, some typical actions could be distinguished at this phase. The most common action was to seek one's way to nature. The contexts often mentioned were forests, lakes or seas. Additional ways of acting were, e.g., staying out of groups of people or just being alone (usually at home) and avoiding communication with others. In some cases, listening music, enjoying art by going to a theatre or to a cinema or visiting an art exhibition at a museum were said to be ways to reach a state of solitude. Some people play a musical instrument by themselves for the same purpose. A few people wrote that religious practises or meditative sessions (in a church or at home) were the key factors in reaching a solitude state.

Surprisingly, a few participants described that they could reach a state of solitude in a bus or by wandering on a crowded street, supposing no one tried to contact them. Interestingly, in this study only a couple of the participants mentioned that they would need or use virtual technology or Internet services in order to get into a state of solitude (e.g., gaming).

The goal of form-of-life analysis in LBD is the identification of design-related *human requirements* (which is different of user requirements). In case of solitude, the most fundamental human requirements were:

- People should be allowed to have an excuse to be 'disconnected' in order to get some time just for themselves for processing their own life, needs and feelings;
- Different ways to experience positive solitude should be taken into account. Even a single person's experiences of solitude may vary (e.g., depending on the context), and the different solitude states should be supported equally;
- Turning to a solitude mode should be possible anytime, i.e., in various situations (contexts) of everyday life;
- Turning back to a 'normal mode' should be guaranteed without having to incur any emotional losses or other mental costs due the solitude mode;

Here we have studied how we can use solitude as a form of life to aid our design thinking. Now that this design basis is identified we can move on to analyse the properties of solitude and create human requirements of it for concept design. After that, a more concrete analysis about the role of technology in fulfilling the human requirements in relation to solitude will be considered.

### *From human* **requirements** *to concept design: The Bazaar of Solitude*

In concept design, technical solutions are innovated to solve problems or to improve possibilities in a defined situation of life. This means that we define people's actions that can be supported with a technology and the role of technology in achieving action goals. Here, the solution ideation is based on the idea of *a Bazaar of Solitude*, which provides us with a cluster or relevant services (Figure 2).

The Bazaar of Solitude is a conceptual model for designing services to support people wanting to live in solitude during short or longer periods of time. Here, it is illustrated what people could do with this service and how they should use it. Also, the technologies which could be used to realize the basic concept should be defined.

There are at least three possible and partly contradictory purposes and uses for technology associated with solitude. Firstly, technology can contribute to the circumstances in a way which enables the state of solitude when needed and regardless of time and place. Secondly, technology may aid people who are in a state of solitude to get into contact with other people without disturbing too much their desire for solitude. Thirdly, technology can enable a complete return from a state of solitude in a way that a full update of the life outside that solitude state is available.

Based on the results of our study, people have different types of solitude states, and these states can be very different from each other. On one hand, solitude may be experienced in the wilderness privacy or with the help of meditation. On the other hand, it can be searched for from, e.g., watching movies or listening to the music. These solitude states are facts that the designer has to examine when creating technological, i.e., human and technical, concepts for a suitable service or application.

To make concrete progress in the design, it is necessary to proceed with three different issues. These are: i) *actions*, i.e. the desired states of solitude, ii) *user interface paradigms*, and iii) *background technologies*. We begin here with actions, and concentrate on three types of actions that would enhance the state of solitude. The types are: meditation, searching for solitude at home, and experiencing the wilderness privacy.

The bazaar of solitude in this case could consist of three alternative solitude concepts: meditation, activities related to listening to favourite music, and dealing with experiences in the wilderness privacy. In common, all of them support immersion to solitude.

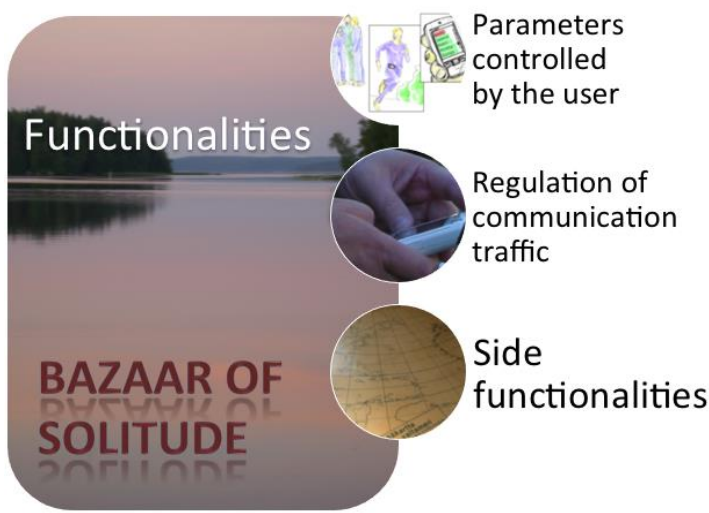

*Figure 2*. *Bazaar of Solitude*.

### *Basic Elements of solitude*

Which are the factors that would prevent or make it difficult to reach the state of solitude in the given contexts? According to the narratives written by our test participants, noise (i.e., noisy traffic, pets, loud music, etc.), unexpected interruptions, contact requests and phone calls may prevent people from getting to their solitude mode. Furthermore, time schedules, duties, hurry and even unpleasant news or places which arouse negative memories, are among them.

The methods for reaching a state of solitude are various and they can sometimes be even culture dependent. Some common features (without going to details here) can be found. They are used here as an example for designing devices and services which might help in meditation. Three basic issues to be tackled are: reducing external noise and disturbing effects, reducing the internal or mental noise of the user, and helping the user to activate a balanced state of mind, which current thoughts and feelings won't disturb.

1. Reducing external disturbances and noise depends very much on the user's physical environment and the context. There are many means for eliminating disturbing physical auditory and visual noise. These means vary from isolation chambers to earphones with active noise compensation and eyeglasses with integrated virtual display. More and more of the disturbances come nowadays from the virtual telecom world. Silent modes, switching off the power and directing calls and messages to answering machines or services are good ways to eliminate this disturbance.

2. Internal mental noise and disturbance comes often from the fact that you feel a necessity to be continuously alert. Your boss may contact you, and your family or other social networks expect you to participate and be socially active. A solitude mode can be built into the telecommunication devices and services. An agent or your personal virtual assistant working at the background could take care of responding and actively participating in communication on your behalf. One would be alerted only in very important or urgent situations.

3. To activate a balanced state of mind is very personal. In many cases, refreshing positive memories is a good way to forget the currently disturbing thoughts. In some forms of meditation, finding a positive setting of your mind is the key. Many times the setting dates back to the childhood. The setting may consist of visions, sounds, places, activity etc. that create a good and balanced feeling. Your private environment, a silent chamber or a virtual device can provide you with relaxing sounds from nature, relaxing speech, your favourite music, images or videos about your favourite places and activities and get you into a positive neutral mode. These all can be based on your personal profile and fitted into your present context.

*Innovative leap*

In the next step, the designers have to make assumptions and decisions about the relevant and suitable technologies that are compatible with the human requirements and can be used to help peoples in their intentions to reach a solitude state. This solution ideation and reflection is a critical stage in Life-Based Design. We may call it the *innovative leap*, where the facts of life and human values are transformed into technical solutions. Innovators' thinking switches from *how things are* (e.g., how people feel and think about something) to *how things should be*, (i.e., how it would be possible to improve the present facts). The term 'leap' describes the fact that this point in innovative thinking is discontinuous.

*Elaboration of selected solutions*

When we think about solitude and technology, mobile technology can naturally be seen as a facilitator and mobile phone as a forerunner in fulfilling users' needs for 'freedom' and 'own space'. Being the device that has met the universal human need for unrestricted mobility, mobile phone has also, controversially, become the device that provides us with the true freedom of *not* having to communicate all the time. While the mobile phone originally increased the freedom to move, it considerably seemed to influence our privacy, as well. It gave us the freedom of not being reachable and the freedom of going away and being on your own (Schmolze 2005). It was this consumer need, the desire to reduce the immediacy of the mobile phone, which also paved the way to the success of text message (SMS).

Another phenomenon that fundamentally changed our concept of time and space was the Internet. No matter whether accessed by smart phones or by desktop computers or laptops, the Internet enables many flexible ways of staying aware without being online all the time. Newspapers may be read and daily news watched or listened to at the time we prefer. Everything we need (including personal emails and social service connections) for keeping us in touch is available for us at any time without having to worry that we might have lost something. This is a fundamental advantage, enabling safe sojourns to solitude.

Lately the bridging between the physical and digital worlds (often referred to with terms such as 'ubiquitous computing', 'pervasive computing' or 'ambient intelligence') has been aimed at connecting everyday things (homes and devices) to each other and to the Internet. The embedded ubiquitous technology can also be used for sensing and adjusting the environment people are living in (e.g., home control systems measuring and regulating heat and lighting). These features of so called 'smart space technology' can be exploited in turning to or returning from one's solitude mode.

If we think that solitude should be available for people at any time and place, the technology should support mobility. Now that people are carrying smart phones everywhere, with efficient access to the Internet all the time, we suggest that smart phones should be harnessed for supporting individual goals of solitude. And for that, it is necessary to specify the user and functional requirements for the required smart phone based applications. For example, the solitude application concept could be turned on by pressing a solitude icon on the screen of a smart phone. The system might propose some alternatives saved into the bazaar of solitude. The selection of a suitable mode of solitude can then be selected and started as an ordinary application on a smart phone.

There are many telecommunication device technologies and networks available today and can be used to support functions needed for different modes of solitude. We won't go deep into technical details in this paper, but some functional principles and high level vision need to be shared. The functionalities can be defined based on the targeted activities in the solitude mode and the corresponding technical architectures in the telecommunication infrastructure. The technologies can be local or global, personal or shared between different user groups, tied to a location or utilizing global telecommunication networks.

Local technologies included in the wireless personal device which can be used to understand user context include different sensors and radio interfaces sensing the location and environment of the user. Microphones, cameras, acceleration meters and positioning technologies, for example, can detect the current activity and environment of the user, whether sleeping, running, walking, working in a silent or noisy environment alone or with friends. Global positioning system (GPS) and near field communication technologies (Blue Tooth, Wireless Local Area Networks etc.) can be used to detect the telecom environment and, for example, the local services available. In the future, sensors detecting the user's medical and mental state, blood pressure, heart beat and even mood are going to be integrated into personal telecom devices.

Global technologies include different services available through the Internet and the intelligence distributed into different server nodes in the telecom infra. The distributed network and servers, cloud-based services, can also be utilized to store data about the user and his behavior. Based on users' personal profiles, including their preferences, by analyzing the statistics of users' behavior, intelligent recommendations can be made. We call this the smart personal assistant or intelligent agent, which serves the user at the background in the solitude mode. Understanding and monitoring the social networks of the user and their dynamics is also important to ensure that the personal assistant makes right selections.

*Personification of the Bazaar of Solitude*

As discussed above, there are various kinds of solitude environments. Hence, we have to make decisions about what kind of solitude state or states the system to be designed should support. There seems to be four kinds of issues relevant here (Figure 3). Firstly, we need user-defined *personal solitude parameters*. With these we refer to different attributes to personify the solitude states that users want. For example, people who use music to get to the solitude state of mind are most likely not interested in the same kind of music. While some people like to listen to sentimental classics, some others may find rock music more fascinating (personification). Thus the users must have an opportunity to modify the service according to their own choices.

Secondly, there will be *user-specified machine-controlled interventions*. For example, people must be able to "fine-tune" the level of external messages. This means that some critical information can be allowed to break the state of solitude, some information may be put on hold to be processed later, and irrelevant information can simply be skipped. This means that the users must have parameters to define the type of messages they regard as relevant and the way the messages will be processed.

Thirdly, there should be numerous *context monitoring functions* to modify any solitude technology that becomes active in the application as a consequence of the selections made by the users. The goal is to make the fine-tuning of the application as easy as possible for the user. These activities monitor various states of users, their contexts and actions, offering improvements to the application by automatically adapting it.

Finally, a solitude technology must also have many *support services and* provide *offerings*. For example, people who like to search for solitude from the wilderness privacy might appreciate the possibility of having effective navigation and map services. They may also wish to gain information about the quality of possible beauty spots and offerings of the nearest Bed & Breakfasts. User's role in these kinds of solutions can be active. The users should be allowed to define the essential modes of these support services for themselves.

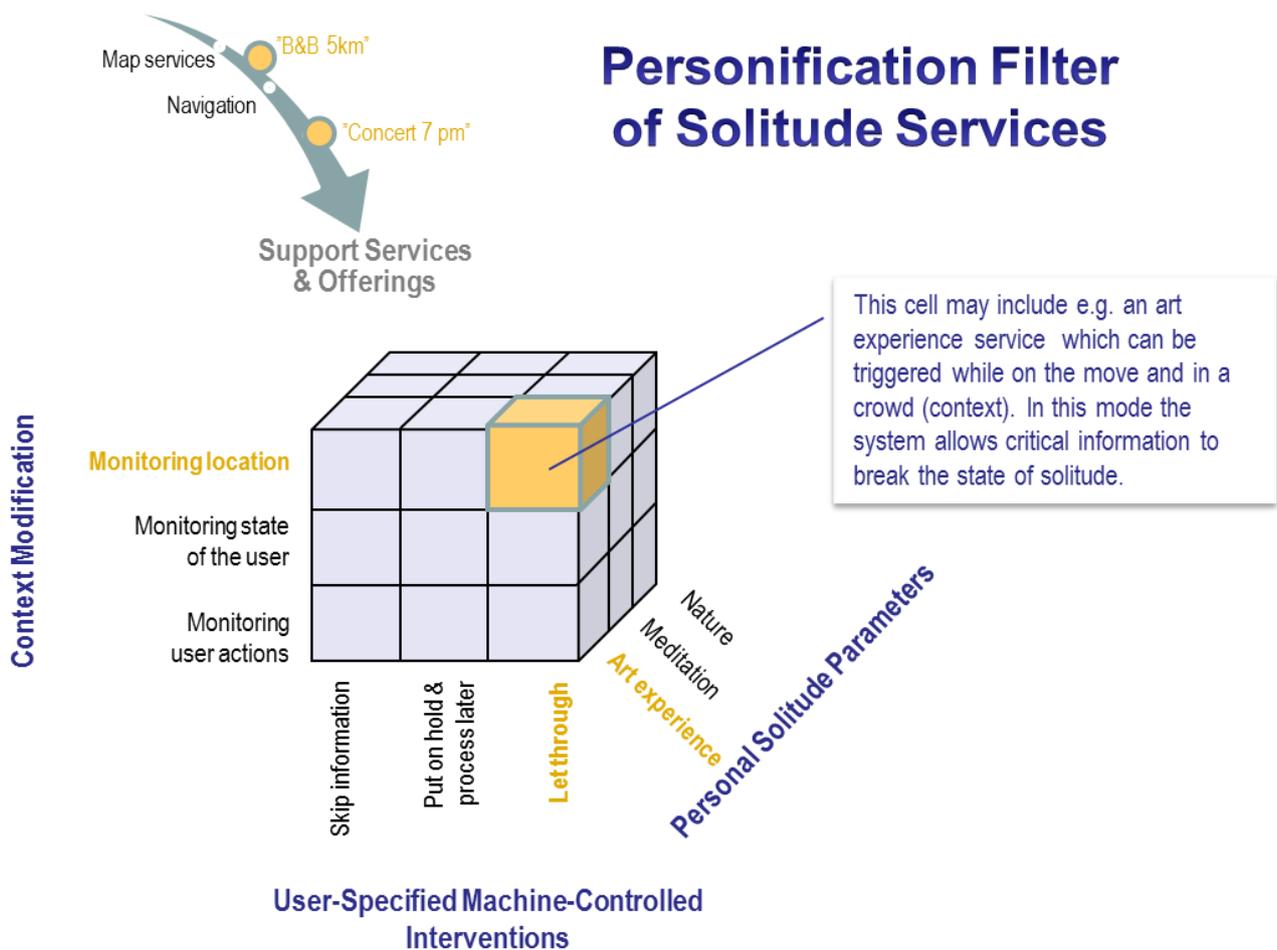

*Figure 3.* Personification filter of solitude services (in practice this matrix is much wider than 3x3x3). Here Personal Solitude Parameters refers to different selections that users can make to fine-tune their device (e.g., "today I'd like to listen to the Queen instead of Bach"). User-Specified Machine-Controlled Interventions refer to different types of system-controlled selections predefined by the users (e.g., "allow calls from my wife but not from my office"). Context Modification refers to different types of contextual information monitored by the device (e.g., geographic location and emotional state of the user). This information is used to provide Support Services (e.g., navigation) and Sense-Making Offerings (e.g., "bed and breakfast 5 km") for the users.

*And back to quality of life*

Fit-for-life design is the phase of the design where the outcome of the concept design should be analysed in terms of its benefits and meaningfulness to the users (Leikas et al. 2013). This is an important phase, as it assesses the impact that the solution has to the quality of people's lives. The design outcome should fit seamlessly to the form of life of the target users and enhance their quality of life.

This analysis is mostly forgotten in the design processes. It is essential that the designers come back to the human requirements defined in the first phase of the process and reflect upon them. Also, with the help of efficient user studies and concept evaluations with the users it is possible to define and remove barriers for a successful fit for life. This phase is closely connected to the final phase of LBD, namely innovation design that introduces the procedure for exporting the design outcome into general use. With the help of a creation of a usage culture and by conveying the message to the users about the benefits from using the service, it is possible to ensure that the service has found its users.

*Discussion*

Life-Based Design is a specific method for developing technology. It differs from traditional human-centred design methods in that it forms not only a conceptual framework based on analysis of human life but also specific conceptualizations and procedures to help designers in their work. Instead of simply concentrating on user-needs, it strives for understanding what people do, what the basic logic of their actions is, and why they do things in some particular way. In doing this, LBD makes the design process more systematic.

Life-Based Design is based on investigating human life (Leikas 2009). Alike, this paper's essential task is to identify some relevant forms of life: a situation in life, a system of values and some system of regular actions people carry out in their lives. In this paper, our focus has been on studying how to use a form of life to aid design thinking. In our example, we have classified solitude as a form of life, and followed the Life-Based Design process in creating a concept for a solitude service, i.e., reaching a state of solitude supported by electronic gadgets. As a form of life, solitude forms here the basis for design, i.e., an idea that can be used to integrate the whole design plan.

We could use paradigmatically similar thinking in solving other types of problems emerging from situations in life. In Life-Based Design, these constitute 'facts of life' (Leikas 2009; Saariluoma and Leikas 2010). Examples of alternative viewpoints to solitude could be, e.g., poverty and aging (Saariluoma, Helfenstein and Maksimainen 2009), although these examples are different from solitude in that they are not consequences of free will in the same sense as solitude is. Nevertheless, when one is put in a certain situation in life or has chosen a certain state of mind, be it even for a very short duration, this particular state defines the relevant technologies that can be used to support or enable this goal to come true.

In concept design, technical solutions are innovated to solve problems or to improve possibilities in a defined situation in life. This means defining the actions of people in a certain context which can be supported by a technology. Here, the solution was based on the idea of the Bazaar of Solitude, which provides us with a cluster of relevant service ideas. In this concept, immediate interaction or usability problems are solved in a coherent manner. All types of solitude-relevant human actions and respective interaction components, such as menus and icons, can be designed in a harmonious and consistent manner. This way the users can reach their action goals easier than in cases of a non-coherent system.

We live in times of dramatic shift in technology design. ICT-technologies have an infinite number of users. Technical devices can be connected to our everyday life in ways that are totally different from what they used to be in the past. It is no longer necessary to have a new tool each time we want to reach new goals by means of technologies. Instead, we can modify the given ICT tools with small application programs so that the same computer or computational tool can be used for numerous different goals in everyday life. Thus, a specific "hardware" can provide us with an unlimited number of possible functions and services. Today, for example, it is possible to keep up contacts by using a variety of electronic devices and applications, such as e-mails, SMS-messages, pictures, video calls and social media.

The recent change in technological environment has made it necessary also for the designers to seek for new ways to understand what the design actually consists of. As ecosystems and a huge number of possible applications give a new direction to the development of technologies and human-technology interaction, we have to start considering areas that are personally and socially challenging for people instead of focusing mainly on technically interesting problems. In other words, we have to start to create functionalities for life – *to design for life*. To make the design really meet people's wishes, the demand for enhancing quality of life with the help of technology has to be taken seriously. When striving towards design to create applications that enable us to enhance our quality of life, the designers have to be aware of people's goals, the human needs behind them and the different ways people are used to fulfil these needs. The Life-Based Design approach was developed (Leikas and Saariluoma 2008; Leikas 2009; Saariluoma and Leikas 2010; Leikas et al. 2013) to meet this demand and to guide in the early phases of interaction design.

As our experience suggests, instead of random ideation, more systematic and converging creativity is needed in human-technology interaction design. Life-Based Design can be effectively used to research and design for different goals of everyday life. Our example shows how it is possible to conceptualize new services quickly by carrying out the Life-Based Design process. Solitude is only one of the many (mental) needs that people have in their life and the task of technology development is to answer these new demands. By structuring the analysis of life and connecting it to design and innovation processes, Life-Based Design provides a good method for constructive innovative thinking, which pursues towards a higher quality of life.